\def\titre{Collective modes and correlations in one-component plasmas}
%\newif\ifversion
%\versionfalse
%\font\eightrm=cmr8
\font\ninerm=cmr9
%\def\versheader{%
%\vbox{%
%\line{\eightrm\titre\hfil\version}%
%\vskip2mm%
%\hrule}}
%\ifversion\headline={\versheader}\fi
\magnification=1200
\headline{\ifnum\pageno=1 \nopagenumbers 
\else
\ifodd\pageno
 \hss\number \pageno 
\else \number\pageno\hss\fi
\fi} 
\overfullrule=0pt
\footline={}
%\footline={\ifnum\pageno=1\nopagenumbers
%\else\centerline{\number \pageno}\fi}
%\nopagenumbers
\font\boldmath=cmbsy10
\textfont2=\boldmath
\font\boldgreek=cmmib10
\textfont9=\boldgreek
\def\rrprime{{\bf r},{\bf r}'}
\def\correl#1#2{<#1 ({\bf r}) #2 ({\bf r}')>^{\rm T}}
\def\dyncorrel#1#2{<#1 ({\bf r},t) #2 ({\bf r}',t')>^{\rm T}}
\def\phiphi{\correl{\Phi}{\Phi}}
\def\dynphiphi{\dyncorrel{\Phi}{\Phi}}
\def\dynsigmasigma{\dyncorrel{\sigma}{\sigma}}
\def\sigmasigma{\correl{\sigma}{\sigma}}
\def\kpara{{k_\parallel}}
\def\kvperp{{{\bf k}_\perp}}
\def\kperp{{k_\perp}}
\def\cphiphi{{\widetilde{C_{\Phi\Phi}}(\kvperp,z,z',\omega)}}
\def\csigmasigma{{\widetilde{C_{\sigma\sigma}}(\kvperp,\omega)}}
\def\sgn{{\rm sgn}}
\def\kbt{{k_{\rm B}T}}

\centerline{{\bf\titre}}
\vskip 10mm
\centerline{G.~T\'ellez\footnote{$^{\dagger}$}{E-mail address~:
tellez@stat.th.u-psud.fr}}
\medskip \centerline{{\it Laboratoire de Physique Th\'eorique et
Hautes Energies}\footnote{*}{Laboratoire associ\'e au Centre National de
la Recherche Scientifique - URA D0063
}} 
\centerline{{\it Universit\'e de Paris-Sud, b\^atiment 211, 91405 Orsay
Cedex, France}}
\bigskip

{
\ninerm
\baselineskip 15pt
\narrower
The static and time-dependent potential and surface charge correlations
in a plasma with a boundary are computed for different shapes of the
boundary. The case of a spheroidal or spherical one-component plasma is
studied in detail because experimental results are available for such
systems.  Also, since there is some knowlegde both experimental
and theoretical about the electrostatic collective modes of these
plasmas, the time-dependent correlations are computed using a method
involving these modes.

\vskip 3mm
\noindent PACs numbers: 05.20.-y, 52.25,Kn, 52.35.Fp, 52.25.Wz, 41.20.Cv

}

\vskip 50mm

LPTHE Orsay 96-39

May 1996

\eject
\baselineskip 20pt
\vskip 5mm
\centerline {\bf I. INTRODUCTION}
\vskip 5mm

The static correlations between charged particles in classical plasmas
at equilibrium have attracted some theoretical attention~[1]. For a
plasma with a boundary, the surface charge correlations are especially
interesting, because they are universal, in that sense that they do not
depend on the detail of the microscopic constitution of the plasma, for
length scales large compared to the microscopic ones~[2] (although these
correlations do depend on the geometry of the boundary).

Experimentally~[3,4], classical plasmas in thermal equilibrium, with a
boundary, have been obtained by confining particles of one sign in a
Penning or Paul trap. These particles may be electrons or ions. In a
Penning trap, a magnetic field along the $z$ axis confines the
particules radially while an electric field due to suitable electrodes
provides the axial confinement. In a Paul trap, the confinement is
provided by an electrostatic potential and a radiofrequency field. In
both cases, one obtains  a system which behaves like a one-component
plasma: a system of particles of one sign immersed in a neutralizing
uniform background (in the experiments, the confining fields play the
role of the background). The particles form a blob of spheroidal shape.
In the case of a Penning trap, the blob performs a rigid rotation around
the $z$ axis, and it is in the corotating frame that it behaves like a
static one-component plasma in a magnetic field; however, if the rotation
frequency is just half the cyclotron frequency (the Brillouin regime),
the system behaves like a one-component plasma without a magnetic
field.

	Electrostatic collective modes of such spheroidal plasmas have
been experimentally observed~[5-8] and theoretically discussed~[9-11].

These experimental findings are an incentive for applying the general
theory~[2] of surface charge correlations to spheroidal plasmas; this
will be done in Section II. Furthermore, since there is some knowledge
about the electrostatic collective modes, it is desirable to check in a
variety of cases that the static charge correlations can be correctly
obtained (on large length scales) as the sum of the contributions from
all the thermally excited collective modes (it was found some time ago
that this approach does work in the very simple special case of a plane
boundary without a magnetic field~[12]); this will be done in Section
III. This method for computing the static correlations also gives the
time-dependent correlations.

\vskip 5mm
\centerline {\bf II. STATIC CORRELATIONS IN A SPHEROIDAL PLASMA}
\vskip 5mm

It should be emphasized that the calculations of this Section are based
on macroscopic electrostatics and give results which are valid only on
macroscopic length scales. Therefore, these calculations cannot account
for the shell or crystal structures which have been observed~[13] for
plasmas in a Penning trap and reproduced by computer simulations~[14].
The present calculations apply to correlations smoothed on microscopic
oscillations. Also, the surface charge density $\sigma$ which will be
considered here should be understood as a microscopic volume charge
density integrated along the normal to the surface and smoothed in
directions parallel to the surface.

As explained in the Introduction, a plasma confined in a Penning trap
behaves, in a rotating frame, like a static spheroidal one-component
plasma submitted to a uniform magnetic field along its axis. We assume
that the Debye length is much smaller than the plasma dimensions: this
condition ensures that the spheroid has a well-defined surface and that
the spheroid size is macroscopic.

It is well known that in classical statistical mechanics a magnetic
field has no effect on the static quantities. Thus we shall compute the
static (equal-time) correlations of a spheroidal one-component plasma in
the absence of a magnetic field. The results will be also applicable to
the case with a magnetic field.

We use the method of Ref.~[2]. Surface charge correlations can be
derived from electric field correlations by considering the
discontinuity of the normal electric field across the surface of the
plasma. The two-point electric potential correlations can be computed on
a macroscopic scale by using linear response theory and macroscopic
electrostatics.  Let us put a test charge $q$ at ${\bf r}$. At some point
${\bf r}'$ the potential change $\delta\Phi({\bf r}')$ due only to
the plasma is related by linear response to the potential correlation at
thermal equilibrium:

$$
\delta\Phi({\bf r}')=-\beta q \phiphi \, ,
\eqno(2.1)
$$ 
where $\beta=(\kbt)^{-1}$ with $T$ the temperature and $k_{\rm B}$ the
Boltzmann constant, and 

\noindent $<AB>^{\rm T}$ means the truncated average $<AB>-<A><B>$.

The potential change $\delta\Phi({\bf r}')$ can be computed using
screening properties of the plasma and macroscopic electrostatic
arguments provided that $|{\bf r}-{\bf r}'|$ is large compared to the
screening length. From now on we shall assume that this condition is
satisfied.

Let us apply this to the particular case of an insulated spheroidal
plasma. We consider a ellipsoid of revolution around the $z$ axis.
Let $2b$ be its axial length and $2a$ its diameter. Let $d^2=b^2-a^2$.
Then $|d|$ is the distance between the foci of the spheroid. If $d^2>0$
then the spheroid is prolate, otherwise it is oblate. We use spheroidal
coordinates $(\xi,\eta, \phi)$ defined by
$$
\eqalign{
x&=[(\xi^2-d^2)(1-\eta^2)]^{1/2}\cos\phi\,,\cr
y&=[(\xi^2-d^2)(1-\eta^2)]^{1/2}\sin\phi\,,\cr
z&=\xi\eta\,.\cr
}
\eqno(2.2)
$$

The boundary between the plasma and the vacuum is then the 
spheroidal surface defined by 
$$
\xi=b\,.\eqno(2.3)
$$ 

Let us consider first the case where ${\bf r}$ and ${\bf r}'$ are inside
the plasma.  Due to the plasma's screening properties the charge $q$ at
${\bf r}$ will be surrounded by a polarization cloud of microscopic
dimensions carrying a charge $-q$ giving to $\delta\Phi({\bf r}')$ a
contribution $-q/|{\bf r}-{\bf r}'|$. Since the plasma is insulated a
charge $+q$ spreads on the surface of the spheroid $\xi=b$. This
surface charge gives to $\delta\Phi({\bf r}')$ another contribution
equal to $q/C$ with the capacitance $C$ given by~[15]
$$ 
C={d\over Q_0(b/d)}\,,
\eqno(2.4)
$$ 
where $Q_0$ is a Legendre function of second kind. Using the total
$\delta\Phi({\bf r}')$ in (2.1) gives, for ${\bf r}$ and ${\bf r}'$
inside the spheroid 
$$
\beta\phiphi={1\over|{\bf r}-{\bf r}'|}-{1\over d}\ Q_0(b/d)
\,.
\eqno(2.5)
$$

If ${\bf r}'$ is outside but ${\bf r}$ is still inside the plasma the
potential change at ${\bf r}'$ created by the surface charge $q$ is
equal to $qQ_0(\xi'/d)/d$. So we have in that case
$$
\beta\phiphi={1\over|{\bf r}-{\bf r}'|}-{1\over d}\ Q_0(\xi'/d)
\,.
\eqno(2.6)
$$

If ${\bf r}$ and ${\bf r}'$ are both outside the plasma the potential
change at ${\bf r}'$ is $q[G(\rrprime)-|{\bf r}-{\bf r'}|^{-1}]$ where
$G(\rrprime)$ is the potential at ${\bf r}'$ when a unit charge is put
at ${\bf r}$ outside the insulated spheroidal conductor. $G$ is given
by~[15]
$$
\eqalign{
G(\rrprime)=&{1\over d} 
\sum_{n=0}^{\infty} (2n+1)
\sum_{m=0}^n \epsilon_m (-1)^m
\left[ {(n-m)!\over(n+m)!} \right]^2
\cos[m(\phi-\phi')]
P_n^m(\eta)P_n^m(\eta')\cr
&\times\left[
-Q_n^m(\xi/d)Q_n^m(\xi'/d)
{P_n^m(b/d)\over Q_n^m(b/d)}
+
\cases{
P_n^m(\xi/d)Q_n^m(\xi'/d), &if $\xi<\xi'$\cr
P_n^m(\xi'/d)Q_n^m(\xi/d), &if $\xi'<\xi$\cr
}
\ \ \right]\cr
&+{Q_0(\xi/d)Q_0(\xi'/d)\over Q_0(b/d)}
\,,
}\eqno(2.7)
$$ 
where $P_n^m$ and $Q_n^m$ are associated Legendre functions of the first
and second kind respectively and $\epsilon_m=2-\delta_{m0}$ is the
Neumann factor. The $P_n^m(\xi/d)Q_n^m(\xi'/d)$ or
$P_n^m(\xi'/d)Q_n^m(\xi/d)$ terms in (2.7) come from an expansion of $|{\bf
r}-{\bf r}'|^{-1}$, and finally the electric potential correlation is
given by
$$
\eqalign{
\beta\phiphi=&
{1\over d} 
\sum_{n=1}^{\infty} (2n+1)
\sum_{m=0}^n \epsilon_m (-1)^m
\left[ {(n-m)!\over(n+m)!} \right]^2
\cos[m(\phi-\phi')]\cr
&\times
P_n^m(\eta)P_n^m(\eta')
Q_n^m(\xi/d)Q_n^m(\xi'/d)
{P_n^m(b/d)\over Q_n^m(b/d)}
\,.
}
\eqno(2.8)
$$
The surface charge correlation is
$$
\sigmasigma={1\over(4\pi)^2}
<(E_n^{\rm out}({\bf r})-E_n^{\rm in}({\bf r})) (E_n^{\rm out}({\bf
r}')-E_n^{\rm in}({\bf r}')) >^{\rm T} 
\,,
\eqno(2.9)
$$ 
where $E_n^{\rm in,(out)}({\bf r})$ denotes the limit of the normal
component of the electric field when ${\bf r}$ approaches the surface
from the inside (outside). Using the expressions (2.5), (2.6) and (2.8)
for the electric potential correlations we find 
$$
\beta\sigmasigma=-{1\over(4\pi)^2}
{b^2-d^2\over \sqrt{(b^2-d^2\eta^2)(b^2-d^2\eta'^2)}}
{\partial^2 G(\rrprime)\over\partial\xi\partial\xi'}
\Bigg|_{{\bf r},{\bf r}'\in{\rm surface}}
\,,
\eqno(2.10)
$$
which finally gives
$$
\eqalign{
\beta\sigmasigma=&
-{1\over (4\pi)^2 d\sqrt{(b^2-d^2\eta^2)(b^2-d^2\eta'^2)}}
\sum_{n=1}^\infty
\sum_{m=0}^n (2n+1)\epsilon_m
\cr
&\times
{(n-m)!\over(n+m)!}
P_n^m(\eta)P_n^m(\eta')
{{Q_n^m}'(b/d)\over Q_n^m(b/d)}
\cos m(\phi-\phi')
\,.
}
\eqno(2.11)
$$
In the case of a globally neutral spheroid, $<\sigma({\bf r})>=0$ and
the truncation sign T may be omitted.

From this last expression we can recover the charge correlation for some
particular geometries. For example if $d$ goes to zero, we have the case
of a spherical plasma. In that limit $b$ becomes the radius of the
sphere and ${Q_n^m}'(b/d)/Q_n^m(b/d)\to -(n+1)d/b$ then equation~(2.11)
becomes
$$ 
\eqalign{
\beta\sigmasigma=&
{1\over (4\pi)^2 b^3}
\sum_{n=1}^\infty
\sum_{m=0}^n (2n+1)(n+1)\epsilon_m
{(n-m)!\over(n+m)!}\cr
&\times
P_n^m(\eta)P_n^m(\eta')
\cos m(\phi-\phi')\, .
}
\eqno(2.12)
$$
The sum can be performed to give the already known result [2,~16] 
$$
\beta\sigmasigma=-{1\over8\pi^2b^3}\left[
{1\over\left(2\sin{\alpha\over2}\right)^3}+
{1\over2}
\right]
\,,
\eqno(2.13)
$$
where $\alpha$ is the angle between ${\bf r}$ and ${\bf r}'$.

Another special case is the cylindrical geometry obtained taking the
limit $b\to\infty$, then $a$ is the radius of the cylinder. In that case
it is interesting to define $k=n/b$.  The sum over $n$ times $b^{-1}$
becomes an integral over $k$, $b/d\sim 1+{a^2\over2b^2}$, $\eta\sim z/b$
and using the asymptotic expansions 
$$
\eqalignno{
Q_n^m(b/d)&\sim(in)^mK_m(ka)\,,&(2.14{\rm a})\cr
P_n^m(\eta)&\sim\sqrt{{2\over
n\pi}}n^m\cos\left[(n-m){\pi\over2}-kz\right]\,,
&(2.14{\rm b})\cr
}
$$
where the $K_m$ are modified Bessel function of the third kind, 
we find
$$
\beta\sigmasigma=
-{1\over8\pi^3}\sum_{m=0}^\infty\epsilon_m
\cos m(\phi-\phi')
\int_0^{+\infty}
{k\over a}{K_m'(ka)\over K_m(ka)}
\cos k(z-z')\, dk
\,.\eqno(2.15)
$$

\vskip 5mm
\centerline {\bf III. COLLECTIVE MODES AND CORRELATIONS}
\vskip 5mm

When the microscopic detail is disregarded, the thermal fluctuations are
expected to be correctly described by the set of collective modes. If
each collective mode $n$ is considered as a harmonic oscillator of
frequency $\omega_n$, the electric potential associated with this mode
is of the form 
$$
\left[\Phi_n({\bf r})e^{-i\omega_nt}+\overline{\Phi_n({\bf
r})}e^{+i\omega_nt}
\right]/\sqrt{2}
\, ,
\eqno(3.1)
$$
with an amplitude of $\Phi_n$ such that the
corresponding average energy be $\kbt$, at temperature $T$. 
Then the
time-displaced potential correlation will be
$$
\dynphiphi={\rm Re}\sum_n
<\overline{\Phi_n ({\bf r})} \Phi_n ({\bf r}')>e^{-i\omega(t-t')}
\, ,
\eqno(3.2)
$$
and the correlation functions of the other functions can be deduced
from (3.2)  

In the general case $t\neq t'$, expression (3.2) is expected to depend
on the magnetic field applied to the plasma. However, the static limit
$t=t'$ should be magnetic field-independent.

The explicit calculation of (3.2) for a spheroidal plasma in a magnetic
field would involve complicated expressions which are not very
illuminating. Therefore, only the special case (A) of a spherical plasma
without a magnetic field is considered here. However, as exercises, we
consider simpler models, with magnetic field, on which it can be
explicitly checked that the static limit $t=t'$ is field-independent.
These models are (B), a plasma along a plane boundary in a magnetic
field normal to the boundary, and (C), a two-dimensional plasma (with
two-dimensional logarithmic Coulomb interactions) in a disk with a
magnetic field normal to the plasma plane.

\vskip 5mm
\centerline {\bf A. Spherical plasma without magnetic field}
\vskip 5mm

We consider in this section a spherical one-component plasma of radius
$R$ composed of particles of mass $m$ and charge $q$ in a uniform
charged background with charge density $-qn_0$, without a magnetic field.
Experimentally this could be achieved in a Penning trap in the Brillouin
regime (in the rotating frame the plasma behaves as an unmagnetized
plasma) or in a Paul trap (as stated above, the confining fields play the
role of the uniform neutralizing background).

We use spherical coordinates $(r,\theta,\phi)$. We compute
the time-dependent correlations which can be seen as the sum of the
contributions from the
different collective modes each one oscillating at its own frequency. It
should be noticed that the time-dependent correlations can also be
computed by a generalization of the linear response method explained in
section II, now using the dynamical linear response theory.

The linearized equations of motion for the electric potential $\Phi$,
the volume charge density $\rho$ and the current density ${\bf j}$
inside the plasma are 
$$
\eqalignno{
\Delta\Phi&=-4\pi\rho\,,&(3.3{\rm a})\cr
{\partial {\bf j}\over\partial t}&=
-{\omega_p^2\over 4\pi}\nabla\Phi\,,
&(3.3{\rm b})\cr
{\partial\rho\over\partial t}&=-\nabla\cdot{\bf j}\,,
&(3.3{\rm c})\cr
}
$$
where $\omega_p=(4\pi q^2n_0/m)^{1/2}$ is the plasma frequency. We look
for a mode of frequency $\omega$.
Manipulating equations (3.3), we find for $\Phi$, inside the plasma, the
equation 
$$
\epsilon\Delta\Phi=0\,,
\eqno(3.4)
$$
where $\epsilon=1-\omega_p^2/\omega^2$. The equation for $\Phi$ outside
the plasma is the usual Laplace equation 
$$
\Delta\Phi=0\,.\eqno(3.5)
$$

The problem has been reduced to the electrostatic problem of a
dielectric filling the sphere of radius $R$. Equations (3.4) and (3.5)
must be supplemented with the boundary conditions
$$
\eqalignno{
&\Phi\to 0 \quad {\rm when} \quad
r\to +\infty\,,&(3.6{\rm a})\cr
&\lim_{r\to R^+} \Phi({\bf r})=\lim_{r\to R^-}\Phi({\bf r})
\,,
&(3.6{\rm b})\cr
&\epsilon\partial_r\Phi(R^-,\theta,\phi)=
\partial_r\Phi(R^+,\theta,\phi)\,.&(3.6{\rm c})\cr
}
$$
Equations (3.4), (3.5) and (3.6) have two types of solutions:

1) Surface modes: for $\epsilon\neq 0$, $\Phi$ satisfies the Laplace
equation inside and outside the plasma. One finds modes (3.1) with
$$
\Phi_{nm}({\bf r})=\cases{
A_{n}r^l Y_n^m(\theta,\phi),&if $r<R$,\cr
A_{n}R^{2l+1}r^{-l-1} Y_n^m(\theta,\phi),&if $r>R$,\cr
}
\eqno(3.7)
$$
and
$$
\omega^2=\omega_{n}^2={n\over 2n+1}\omega_p^2\,,
\eqno(3.8)
$$
where $n$ and $m$ are integers ($n>0$ and $|m|\leq n$) and
$Y_n^m$ are the spherical harmonics.

Equating the average potential energy of this mode to $k_{\rm B}T/2$
gives the average squared amplitude 
$$
\beta<|A_{n}|^2>={4\pi\over(2n+1)R^{2n+1}}
\,.\eqno(3.9)
$$ 
Using the time-displaced analog of (2.9), one find that
the surface modes contribute to the time-dependent surface
charge correlation $\dynsigmasigma$ a term
$$
\eqalign{
\sum_{nm}
{(2n+1)^2\over(4\pi)^2}R^{2n-2}&<|A_{n}|^2>
Y_n^m(\theta,\phi)\overline{Y_n^m(\theta',\phi')}
\cos \omega_{n}(t-t')
=\cr
&k_{\rm B}T\sum_{nm}{2n+1\over4\pi R^3}
Y_n^m(\theta,\phi)\overline{Y_n^m(\theta',\phi')}
\cos \omega_{n}(t-t')
\,.}
\eqno(3.10)
$$

2) Volume modes: when $\epsilon=0$, then $\omega^2=\omega_p^2$. There is now
an infinite number of modes for each ($n$,$m$): $\Phi^{\rm out}=0$ and any
$\Phi^{\rm in}=f(r)Y_n^m(\theta,\phi)$ with $f(R)=0$ is acceptable.
However, a complete basis for the $\Phi^{\rm in}$ can be choosen as the
eigenfunctions of the Laplacian with Dirichlet boundary conditions:
$$
\Phi^{\rm in}({\bf r})=\sum_\gamma a_\gamma f_\gamma({\bf r})\,,
\eqno(3.11)
$$ 
with $\Delta f_\gamma=\lambda_\gamma f_\gamma$,
$f_\gamma(R,\theta,\phi)=0$ and $\int |f_\gamma({\bf r})|^2\,d{\bf r} = 1$.

Equating the average potential energy of each mode to $k_{\rm
B}T/2$, we find 

\noindent $\beta<|a_\gamma|^2>=-4\pi/\lambda_\gamma$. 
So, the contribution of the volume modes to $\dynphiphi$ is
$$
\eqalign{
\sum_\gamma <|a_\gamma|^2>
f_\gamma({\bf r})\overline{f_\gamma({\bf r}')}
\cos\omega_p(t-t')
&=
-4\pi k_{\rm B}T \sum_\gamma \lambda_\gamma^{-1} 
f_\gamma({\bf r})\overline{f_\gamma({\bf r}')}
\cos\omega_p(t-t')\cr
&=
-4\pi k_{\rm B}T G_{\rm D}({\bf r},{\bf r}')
\cos\omega_p(t-t')
\,,
}
\eqno(3.12)
$$
where $G_{\rm D}$ is the Green function of the Laplacian with Dirichlet
boundary conditions on the sphere:
$$
G_{\rm D}({\bf r},{\bf r}')={1\over 4\pi}
\left[
\left| {r\over R}{\bf r}'-{R\over r}{\bf r} \right|^{-1}
-|{\bf r}-{\bf r}'|^{-1}
\right]\,.
\eqno(3.13)
$$
And finally the contribution of the volume modes to $\dynsigmasigma$ is
found to be
$$
{k_{\rm B}T\over 8\pi^2}
{1\over\left(2R\sin(\alpha/2)\right)^3}
\cos\omega_p(t-t')\,,
\eqno(3.14)
$$
where $\alpha$ is the angle between ${\bf r}$ and ${\bf r}'$.
Putting (3.10) and (3.14) together 
$$
\eqalign{
\beta\dynsigmasigma=&
\sum_{n,m}
{2n+1\over4\pi R^3}
Y_n^m(\theta,\phi)\overline{Y_n^m(\theta',\phi')}
\cos \omega_{n}(t-t')
\cr
&+
{1\over 8\pi^2}
{1\over\left(2R\sin(\alpha/2)\right)^3}
\cos\omega_p(t-t')
\,.}
\eqno(3.15)
$$
For $t=t'$ the sum in (3.15) can be performed and we recover the static
result (2.13).

\vskip 5mm
\centerline {\bf B. Plasma in a half space with a magnetic field}
\vskip 5mm

Let us consider now a one-component plasma filling the half space $z<0$
with a uniform magnetic field in the $z$ direction,
${\bf B}=B{\bf\hat{z}}$. Sum rules for the time-dependent correlations
have been obtained for this case using the dynamical linear response
theory~[17]. Here we use the collective mode method to find
expressions for the correlations valid macroscopically.  This collective
mode method has been used in the case ${\bf B}=0$ in~[12], here
we extend it to the case ${\bf B}\neq0$. With the same
notation as in section III-A, the linearized equations of motion now are
$$
\eqalignno{
\Delta\Phi&=-4\pi\rho\,,&(3.16{\rm a})\cr
{\partial {\bf j}\over\partial t}&=
-{\omega_p^2\over 4\pi}\nabla\Phi+\Omega{\bf j}\wedge{\bf\hat{z}}
\,,&(3.16{\rm b})\cr
{\partial\rho\over\partial t}&=-\nabla\cdot{\bf j}
\,,&(3.16{\rm c})\cr
}
$$
where $\Omega=qB/m$ is the cyclotron frequency.

As above we look for a mode with frequency $\omega$. From
equations (3.16) we find for $\Phi$ inside the plasma
$$
\nabla\cdot{\boldmath \epsilon}\nabla\Phi=0
\,,
\eqno(3.17)
$$
where $\epsilon$ is now the plasma dielectric tensor defined in Cartesian
coordinates by
$$
\epsilon=\pmatrix{
\epsilon_1&-i\epsilon_2&0\cr
i\epsilon_2&\epsilon_1&0\cr
0&0&\epsilon_3\cr
}\,,\eqno(3.18)
$$
with $\epsilon_1=1-\omega_p^2/(\omega^2-\Omega^2)$,
$\epsilon_2=\Omega\omega_p^2/[\omega(\omega^2-\Omega^2)]$, and
$\epsilon_3=1-\omega_p^2/\omega^2$.
As to $\Phi$ outside the plasma ($z>0$), it obeys
$$\Delta\Phi=0\,.\eqno(3.19)$$

We now have the problem of an
anisotropic dielectric filling the $z<0$ half space. The boundary
conditions are  
$$
\eqalignno{
&\Phi\to 0 \quad {\rm when} \quad
z\to +\infty\,,&(3.20{\rm a})\cr
&\Phi(x,y,0^-)=\Phi(x,y,0^+)\,,&(3.20{\rm b})\cr
&\epsilon_3\partial_z\Phi(x,y,0^-)=\partial_z\Phi(x,y,0^+)
\,.&(3.20{\rm c})\cr
}
$$
Let us first consider the case $\epsilon_1/\epsilon_3<0$ which give
modes with a frequency in the ranges
$0<|\omega|<{\rm min}(\omega_p,\Omega)$ called magnetized plasma
modes and ${\rm max}(\omega_p,\Omega)<|\omega|<\Omega_u=(\omega_p^2+\Omega^2)^{1/2}$
called upper hybrid modes.

We look for a mode of the form 
$$
\Phi_{\bf k}({\bf r})=
\cases{
(Ae^{i\kpara z}+Be^{-i\kpara z})e^{i\kvperp\cdot{\bf
r}_\perp},&for $z<0$,\cr
(A+B)e^{-\kperp z}
e^{i\kvperp\cdot{\bf r}_\perp},&for $z>0$,\cr
}
\eqno(3.21)
$$
where ${\bf k}=\kvperp+\kpara {\bf\hat{z}}$, with $\kvperp$ in the
$xy$ plane, $\kperp=|\kvperp|$ and ${\bf r}_\perp=(x,y,0)$. Laplace
equation (3.19) is satisfied for $z>0$ and equation (3.17) gives the
dispersion relation
$$
\left(1-{\omega_p^2\over\omega^2}\right)\kpara^2+
\left(1-{\omega_p^2\over\omega^2-\Omega^2}\right)\kperp^2=0\,.
\eqno(3.22)
$$
Solving this equation we find two modes:
one upper hydrid mode with frequency $\omega_+$ and 
one magnetized plasma mode with frequency $\omega_-$:
$$
\omega_\pm^2={1\over 2}\left[
\omega_p^2+\Omega^2\pm
\sqrt{(\omega^2+\Omega^2)^2-4\omega_p^2\Omega^2{\kpara^2\over\kperp^2+\kpara^2}
}\right]
\,.
\eqno(3.23)
$$

Equation (3.20c) gives a relation between the incident and reflected
amplitudes $A$ and $B$
$$
{A\over B}=
{-1-\epsilon_1\epsilon_3+2i\epsilon_3\sqrt{-\epsilon_1/\epsilon_3}
\over 1-\epsilon_1\epsilon_3}\,.
\eqno(3.24)
$$
It should be noted that $|A/B|=1$: there is total reflection on the
surface $z=0$.

The energy of this mode is a quadratic form of
$\partial\Phi\over\partial t$ and $\Phi$. In the particular case $B=0$
this quadratic form is diagonal [12], which means that potential and
kinetic energy have equal averages and by the energy equipartition
theorem each average is $k_{\rm B}T/2$ (this was the case of section
III-A). If $B\neq0$, the potential and kinetic average energies are
different (take for example the simple case of a single charged
particule with a circular trajectory in the plane normal to a magnetic
field when the whole energy is kinetical).  But we can still use
the energy equipartition theorem and say that the total average energy of each
mode is equal to $k_{\rm B}T$.

The total average energy for a large volume $V$ of plasma is
found to be
$$
<E>=V{<|A|^2>\kperp^2\over2\pi}
{\Omega^2\omega_p^2(2\omega^2-\Omega^2-\omega_p^2)
\over(\omega^2-\Omega^2)^2(\omega^2-\omega^2_p)}
\,.
\eqno(3.25)
$$
Equating $<E>$ to $k_{\rm B}T$ gives the average squared amplitude
$<|A|^2>$ of the mode.

Finally these modes will give a contribution to the electric potential
time-dependent correlation $\dynphiphi$ equal to 
$$
V\int {d^3{\bf k}\over(2\pi)^3}
\left[<
\overline{\Phi_{\bf k,+}({\bf
r})}\Phi_{\bf k,+}({\bf r}')
>
\cos\omega_+(t-t')
+
<
\overline{\Phi_{\bf k,-}({\bf
r})}\Phi_{\bf k,-}({\bf r}')
>
\cos\omega_-(t-t')
\right]
\,.
\eqno(3.26)
$$
where $\Phi_{\bf k,+}$ is the electric potential for the upper hybrid
mode and $\Phi_{\bf k,-}$ the potential for the magnetized mode.

To this contribution we must add the one from the possible modes in the
range ${\rm min}(\omega_p,\Omega)<|\omega|<{\rm max}(\omega_p,\Omega)$,
called evanescent modes. In this case $\epsilon_1/\epsilon_3>0$ and we
look for a solution of the form 
$$
\Phi_\kvperp({\bf r})=\cases{
Ce^{i\kvperp\cdot{\bf r}_\perp+\sqrt{\epsilon_1/\epsilon_3}\kperp z},
&if $z<0$,\cr
Ce^{i\kvperp\cdot{\bf r}_\perp-\kperp z},
&if $z>0$,\cr
}
\eqno(3.27)
$$ 
This form satisfies equations (3.17), (3.19) and (3.20a,b). Equation
(3.20c) implies $\epsilon_3<0$ and $\epsilon_1\epsilon_3=1$. This means
that we have an evanescent mode only if $\Omega<\omega_p$ with frequency
given by $\omega^2=\omega_e^2=(\Omega^2+\omega_p^2)/2$. To compute
$<|C|^2>$ we proceed to compute the average total energy of the mode and
equate it to $k_{\rm B}T$. In this case the energy is proportional to
the surface $S$ of the boundary between the plasma and the vacuum. We
finally find 
$$
\beta<|C|^2>={2\pi\over S\kperp}{\omega_p^2-\Omega^2\over\omega_p^2}
\,.
\eqno(3.28)
$$
This mode adds a contribution to $\dynphiphi$ equal to
$$
S\int {d^2\kvperp\over(2\pi)^2}
<\overline{\Phi_\kvperp({\bf r})}
\Phi_\kvperp({\bf r}')>
\cos\omega_e(t-t')
\,.
\eqno(3.29)
$$

In equation (3.26) it is convenient to make a change of variable in the
integral over $\kpara$ and have an integral over $\omega$. In this way
we can express the electric potential correlation $\dynphiphi$ in terms
of its Fourier transform $\cphiphi$ with respect to time and the $x$ and
$y$ coordinates: 
$$
\dynphiphi=\int {d^2 \kvperp\over(2\pi)^2}
\int_{-\infty}^{+\infty} {d\omega}
\cphiphi e^{-i\omega(t-t')+i\kvperp\cdot({\bf r}_\perp-{\bf r}'_\perp)}
\,.
\eqno(3.30)
$$

$\cphiphi=0$ if $|\omega|\in]{\rm min}(\omega_p,\Omega),{\rm
max}(\omega_p,\Omega)[\cup]\Omega_u,+\infty[$ except maybe at
$\omega=\pm\omega_e$ if $\Omega<\omega_p$. And if $\omega$ is not in that
range

\noindent --if $z>0$ and $z'>0$:
$$
\beta\cphiphi=
-{4\epsilon_3\over\omega\kperp}{(-\epsilon_1/\epsilon_3)^{1/2}\over1-\epsilon_1\epsilon_3}e^{-\kperp(z+z')}
\,,
\eqno(3.31{\rm a})
$$
\noindent --if $z<0$ and $z'<0$:
$$
\eqalign{
\beta\cphiphi=&
-{2\over\omega\kperp\epsilon_3(-\epsilon_1/\epsilon_3)^{1/2}}\Bigg[
\cos\left[\kperp(-\epsilon_1/\epsilon_3)^{1/2}(z-z')\right]\cr
&-{1\over1-\epsilon_1\epsilon_3}
\Bigg((1+\epsilon_1\epsilon_3)\cos\left[\kperp(-\epsilon_1/\epsilon_3)^{1/2}
(z+z')\right]
\cr
&+2\epsilon_3 (-\epsilon_1/\epsilon_3)^{1/2}
\sin\left[\kperp(-\epsilon_1/\epsilon_3)^{1/2}(z+z')\right]\Bigg)\Bigg]
\,,
}
\eqno(3.31{\rm b})
$$
\noindent --if $z<0$ and $z'>0$:
$$
\eqalign{
\beta\cphiphi=&
-{4\over\omega\kperp}{e^{-\kperp z'}\over 1
-\epsilon_1\epsilon_3}
\Bigg(
-\sin\left[(-\epsilon_1/\epsilon_3)^{1/2}\kperp z\right] 
\cr
&+\epsilon_3(-\epsilon_1/\epsilon_3)^{1/2} \cos\left[
(-\epsilon_1/\epsilon_3)^{1/2}\kperp z\right]\Bigg)
\,.
}
\eqno(3.31{\rm c})
$$

In the case $\Omega<\omega_p$ it must be added to $\beta\cphiphi$ the term
corresponding to the evanescent mode:
$$
{\pi\over\kperp}{\omega_p^2-\Omega^2\over\omega_p^2}
(\delta(\omega-\omega_e)+\delta(\omega+\omega_e))
\cases{
\exp{-\kperp(z+z')}&if $z>0$ and $z'>0$,\cr
\exp{\kperp{\omega_p^2+\Omega^2\over\omega_p^2-\Omega^2}
(z+z')}&if $z<0$ and $z'<0$,\cr
\exp{\kperp({\omega_p^2+\Omega^2\over\omega_p^2-\Omega^2}z-z')}&if $z<0$ and $z'>0$.\cr
}\eqno(3.32)
$$

From these expressions we can compute the surface charge correlation,
using equation (2.9). The surface charge correlation Fourier
transform is found to be
$$
\eqalign{
\beta\csigmasigma=&
-{4\kperp(1-\epsilon_3)^2\over\omega\epsilon_3(-\epsilon_1/\epsilon_3)^{1/2}
(1-\epsilon_1\epsilon_3)}\cr
&
+{\kperp\over4\pi}{\omega_p^2\over\omega_p^2-\Omega^2}
(\delta(\omega-\omega_e)+\delta(\omega+\omega_e))
\,.}
\eqno(3.33)
$$
The last term is to be included only when $\Omega<\omega_p$.

Now, let us briefly show how these results can be obtained using
dynamical linear response. If we put at $z=z'$ an oscillating charge density 
$\delta(z-z')\exp i(\kvperp\cdot{\bf r}_\perp-\omega t)$, the electric
potential change at $z$ is $\chi(\kvperp,z,z',\omega)\exp
i(\kvperp\cdot{\bf r}_\perp-\omega t)$. The response function $\chi$ is
related to the Fourier transform $\cphiphi$ of the time-dependent
correlation $\dynphiphi$ in the non-perturbed system by the
fluctuation-dissipation theorem:
$$
\beta\cphiphi=-{1\over\pi\omega}{\rm Im}\, \chi(\kvperp,z,z',\omega)
\, ,
\eqno(3.34)
$$ 
and $\chi(\kvperp,z,z',\omega)=
\Psi(\kvperp,z,z',\omega)-2\pi\exp[-\kperp|z-z'|]/\kperp$
where $\Psi$ is the total electric potential, due to the plasma and the
external charge, solution of
$$
\nabla\cdot\epsilon\nabla \left[\Psi(\kvperp,z,z',\omega)
e^{i(\kvperp\cdot{\bf r}_\perp)}\right]=
-4\pi
\delta(z-z')\exp i(\kvperp\cdot{\bf r}_\perp)
\,,\eqno(3.35)
$$
where $\epsilon$ is the dielectric tensor given by (3.18) if $z<0$ or
equal to $1$ if $z>0$, and the boundary conditions (3.20).

Solving equation (3.35) gives for $\chi$

\noindent --if $z<0$ and $z'<0$
$$
\eqalign{
\chi(\kvperp,z,z',\omega)={2\pi\over\kperp}
\Bigg[&
{1\over\epsilon_3}\sqrt{{\epsilon_3\over\epsilon_1}}
\exp\left[-\kperp\sqrt{{\epsilon_1\over\epsilon_3}}|z-z'|\right]
\cr
&+{1-\epsilon_3^{-1}\sqrt{\epsilon_3/\epsilon_1}\over
1+\epsilon_3\sqrt{\epsilon_1/\epsilon_3}}
\exp\left[-\kperp\sqrt{{\epsilon_1\over\epsilon_3}}(z+z')\right]
\cr
&-\exp[-\kperp|z-z'|]
\Bigg]
\,,
}
\eqno(3.36{\rm a})
$$

\noindent --if $z>0$ and $z'<0$
$$
\eqalign{
\chi(\kvperp,z,z',\omega)={2\pi\over\kperp}
\Bigg[&
{2\over1+\epsilon_3\sqrt{\epsilon_1/\epsilon_3}}
\exp\left[-\kperp z+\sqrt{{\epsilon_1\over\epsilon_3}}\kperp z'
\right]
\cr
&-\exp[-\kperp|z-z'|]
\Bigg]
\,,
}
\eqno(3.36{\rm b})
$$

\noindent --if $z>0$ and $z'>0$
$$
\chi(\kvperp,z,z',\omega)={2\pi\over\kperp}
{1-\epsilon_3\sqrt{\epsilon_1/\epsilon_3}\over
1+\epsilon_3\sqrt{\epsilon_1/\epsilon_3}}
\exp\left[-\kperp(z+z')\right]
\,.
\eqno(3.36{\rm c})
$$

From these expressions it is easy to verify that equation (3.34) leads to
(3.31) and (3.32) and therefore both methods give the same results.
Furthermore, for $t=t'$, equation (3.34) and
Kramers-Kronig relation 
$$
\pi{\rm Re}\,\chi(\kvperp,z,z',0)={\rm P}\int_{-\infty}^{+\infty}
{{\rm Im}\,\chi(\kvperp,z,z',\omega)\over\omega}\, d\omega
\,,
\eqno(3.37)
$$
give the well-known static correlation~[2] 
$$
\beta\phiphi=\cases{
|{\bf r}-{\bf r'}|^{-1},&if $z<0$ or $z'<0$,
\cr
\left[
|{\bf r}_\perp-{{\bf r}_\perp}'|^2+(z+z')^2
\right]^{-1/2},
&if $z>0$ and $z'>0$,\cr
}
\eqno(3.38)
$$
which is, as expected, independent of the
presence of the magnetic field.

\vskip 5mm
\centerline {\bf C. Plasma in a disk with a magnetic field}
\vskip 5mm

Here we consider the model of a two-dimensional a one-component
plasma in a disk of radius $R$ with a magnetic field normal to the plane
where the disk lies. The particles interact through the two-dimensional
Coulomb potential $-\ln r$.
We look for modes with frequency
$\omega$. The dielectric formalism is also valid in this case. 
Only the definition of the plasma frequency is slightly changed:
$\omega_p=(2\pi q^2n_0/m)^{1/2}$. The equation for the potential $\Phi$
inside the disk is
$$
\epsilon_1\Delta \Phi = 0
\,.\eqno(3.39)
$$
while outside the disk it is the usual Laplace equation. In polar
coordinates $(r,\theta)$ the boundary conditions are
$$
\eqalignno{
&\Phi\to 0 \quad {\rm when} \quad
r\to +\infty\,,&(3.40{\rm a})\cr
&\lim_{r\to R^+} \Phi({\bf r})=\lim_{r\to R^-}\Phi({\bf r})
\,,&(3.40{\rm b})\cr
&\epsilon_1{\partial\Phi(R^-,\theta)\over\partial r}
-i\epsilon_2{\partial\Phi(R^-,\theta)\over r\partial\theta}
=
{\partial\Phi(R^+,\theta)\over\partial r}
\,.&(3.40{\rm c})\cr
}
$$
As in section III-A, there are two types of solutions:

1) If $\epsilon_1\neq 0$, $\Phi$ satisfies the Laplace equation inside and
outside the disk
$$
\Phi({\bf r})=\cases{
A_{m}r^{|m|} e^{im\theta},&if $r<R$,\cr
A_{m}R^{2|m|}r^{-|m|} e^{im\theta},&if $r>R$,\cr
}
\eqno(3.41)
$$
and there are two possibles frequencies for each integer $m$ ($m\neq0$),
$\omega=\sgn(m)\omega_\pm$, where $\sgn(m)$ denotes de sign of $m$ and
$$
\omega_{\pm}=\left(
-\Omega\pm\sqrt{\Omega^2+2\omega_p^2}
\right)/2
\,,
\eqno(3.42)
$$

Equating the total average energy of the mode to $k_{\rm B}T$ 
gives
$$
\beta <|A_m|^2>={1\over |m|R^{2|m|}}{\omega'\over
\omega'-\omega}\,,
\eqno(3.43)
$$
where $\omega$ is the frequency of the mode and $\omega'$ the other
root of (3.42). The contribution from these modes to
$\dynsigmasigma$ is 
$$
\kbt 
\sum_{m}
{|m|e^{im(\theta-\theta')}\over (\pi R)^2 (\omega_+-\omega_-)}
\left[
\omega_+e^{-i\sgn(m)\omega_-(t-t')}-
\omega_-e^{-i\sgn(m)\omega_+(t-t')}
\right]\, .
\eqno(3.44)
$$

2) If $\epsilon_1=0$, $\omega^2=\omega_p^2+\Omega^2$ and any $\Phi$
satisfies equation (3.39) inside the disk. However, writting $\Phi$ as a
Fourier series in $e^{im\theta}$, boundary conditions (3.40) and Laplace
equation for $\Phi$ outside the disk implies that $\Phi=0$ outside the
disk. The situation is similar to the one in section III-A~2). $\Phi$ can
be written in the base of the eigenfunctions of the Laplacian with
Dirichlet conditions on the boundary. Then equating the average total energy
of each mode to $\kbt$ and following the calculations from section~III-A~2) 
we find the contribution from these modes to $\dynphiphi$:
$$
-2\pi\kbt{\omega_p^2\over\Omega^2+\omega_p^2}G_{\rm D}({\bf r},{\bf r'})
\cos\left[(\omega_p^2+\Omega^2)^{1/2}(t-t')\right]\, ,
\eqno(3.45)
$$
where $G_{\rm D}$ now is the Green function of the Laplacian with Dirichlet
boundary conditions on the disk:
$$
G_{\rm D}({\bf r},{\bf r}')={1\over 2\pi}
\left[
\ln|{\bf r}-{\bf r}'|
-
\ln\left|{\bf r}-(R/r')^2{\bf r}'\right|
\right]\, .
\eqno(3.46)
$$
In two dimensions there is another mode corresponding to $\omega=0$. The
dielectric formalism does not applies here ($\epsilon_2$ diverges when
$\omega=0$). Dealing directly with the equations of motion, we can show
that $\Phi=0$ outside the disk. Then the contribution of this mode can be
computed in the same way as in the case $\epsilon_1=0$. The
contribution to $\dynphiphi$ is found to be
$$
-2\pi\kbt{\Omega^2\over\Omega^2+\omega_p^2}G_{\rm D}({\bf r},{\bf r'})
\, .\eqno(3.47)
$$
Putting together all contributions gives
$$
\eqalign{
\beta\dynsigmasigma
&=
\sum_m
{|m|e^{im(\theta-\theta')}\over (\pi R)^2 (\omega_+-\omega_-)}
\left[
\omega_+e^{-i\sgn(m)\omega_-(t-t')}-
\omega_-e^{-i\sgn(m)\omega_+(t-t')}
\right]
\cr
&+
{\Omega^2+\omega_p^2 \cos\left[(\omega_p^2+\Omega^2)^{1/2}(t-t')\right]
\over\Omega^2+\omega_p^2}
{1\over8\left(\pi R\sin{\theta\over2}\right)^2}\, .
}
\eqno(3.48)
$$
For $t=t'$, the sum in equation (3.48) can be performed and
the static result~[2], independent of the magnetic field ${\bf B}$ is
recovered 
$$
\beta\sigmasigma
=-
{1\over8\left(\pi R\sin{\theta\over2}\right)^2}\, .
\eqno(3.49)
$$

\vskip 5mm
\centerline {\bf CONCLUSION}
\vskip 5mm

Since one-component plasmas have been obtained experimentally, it would
be interesting if the static or dynamical correlations could be measured
and compared to the expressions obtained here. It should be noted that
our calculations give only the dominant term of the expansion of the
correlations in powers of the distance $r$ between the points. It has
been shown that there are other terms in the asymptotic expansion~[18]
of the time-dependent correlations (behaving like $r^{-6}$ for the
potential correlations in an infinite plasma). These algebraic
corrections vanish in the static case.

\vskip 5mm
\centerline {\bf ACKNOWLEDGMENTS}
\vskip 5mm

The author wish to thank B.~Jancovici for useful discussions and
for reading the manuscript.

\vfill \supereject 
\noindent {\bf REFERENCES} \par
\vskip 5mm 

\item{[1]} For a review of previous work, see Ph.~A.~Martin,
Rev.~Mod.~Phys. {\bf 60}, 1075 (1988).

\item{[2]}~B.~Jancovici, J.~Stat.~Phys. {\bf 80} 445 (1995).

\item{[3]}~J.~S.~deGrassie and J.~H.~Malmberg, Phys.~Rev.~Lett. {\bf
39}, 1077 (1977).

\item{[4]} L.~R.~Brewer, J.~D.~Prestage, J.~J.~Bollinger, W.~M.~Itano,
D.~J.~Larson, and D.~J.~Wineland, Phys.~Rev.~A {\bf 38}, 859 (1988).

\item{[5]}~J.~J.~Bollinger, D.~J.~Heinzen, F.~L.~Moore, W.~M.~Itano,
D.~J.~Wineland and D.~H.~E.~Dubin, Phys.~Rev.~A {\bf 48}, 525 (1993).

\item{[6]} C.~S.~Weiner, J.~J.~Bollinger, F.~L.~Moore, and
D.~J.~Wineland, Phys.~Rev.~A {\bf 49}, 3842 (1994).

\item{[7]} M.~D.~Tinkle, R.~G.~Greaves, C.~M.~Surko, R.~L.~Spencer, and
G.~W.~Mason, Phys.~rev.~Lett. {\bf 72}, 352 (1994).

\item{[8]} R.~G.~Greaves, M.~D.~Tinkle, and C.~M.~Surko,
Phys.~Rev.~Lett. {\bf 74}, 90 (1995).

\item{[9]}~D.~H.~E.~Dubin, Phys.~Rev.~Lett.~{\bf 66},2076~(1991).

\item{[10]}~D.~H.~E.~Dubin, Phys.~Fluids B {\bf 5}, 295 (1993).

\item{[11]}~D.~H.~E.~Dubin, summited to Phys.~Rev.~E.

\item{[12]}~B.~Jancovici, J.~Stat.~Phys. {\bf 39}, 427 (1985).

\item{[13]} J.~N.~Tan, J.~J.~Bollinger, B.~Jelenkovic, and
D.~J.~Wineland, Phys.~Rev.~Lett. {\bf 75}, 4198 (1995).

\item{[14]} D.~H.~E.~Dubin and T.~M.~O'Neil, Phys.~Rev.~Lett. {\bf 60},
511 (1988).

%\item{[2]}~S.~Gilbert, J.~Bollinger, and D.~Wineland, Phys.~Rev.~Lett.
%{\bf 60}, 2022 (1988).
%
%\item{[5]}~D.~J.~Heinzen, J.~J.~Bollinger, F.~L.~Moore, W.~M.~Itano and
%D.~J.~Wineland, Phys.~Rev.~Lett. {\bf 66}, 2080 (1991).
%
%\item{[6]}~J.~H.~Malmberg and T.~M.~O'Neil, Phys.~Rev.~Lett. {\bf 39},
%1333 (1977).

\item{[15]}~P.~M.~Morse and H.~Feshbach, {\it Methods of Theoretical
Physics} (McGraw Hill 1953), Chap.~10.

\item{[16]}~Ph.~Choquard, B.~Piller, R.~Rentsch, and P.~Vieillefosse,
J.~Stat.~Phys. {\bf 55}, 1185 (1989).

\item{[17]}~B.~Jancovici, N.~Macris and Ph.~A.~Martin, J.~Stat.~Phys.
{\bf 47}, 229 (1986).

\item{[18]}~A.~Alastuey and Ph.~A.~Martin, Europhys.~Lett.~{\bf 6} (5),
385 (1988).

\end